\documentclass[pra,twocolumn,superscriptaddress,showpacs]{revtex4}
\usepackage{amsmath,amssymb,amsthm}
\usepackage{graphics}
\usepackage{bbm}

\newcounter{thmcounter}
\newtheorem{lemma}[thmcounter]{Lemma}
\newtheorem{observation}[thmcounter]{Observation}
\newtheorem{corollary}[thmcounter]{Corollary}

\newcommand{\be}{\begin{eqnarray}}
\newcommand{\ee}{\end{eqnarray}}
\newcommand{\bee}{\begin{equation}}
\newcommand{\eee}{\end{equation}}
\newcommand{\one}{\mathbbm{1}}
\newcommand{\Tr}[1]{\mathrm{Tr}\left[ {#1} \right]}
\newcommand{\ket}[1]{|{#1}\rangle}
\newcommand{\bra}[1]{\langle{#1}|}
\newcommand{\braket}[2]{\langle{#1}|{#2}\rangle}
\newcommand{\ketbrad}[1]{\left|{#1}\rangle\!\langle{#1}\right|}
\newcommand{\ketbra}[2]{\left|{#1}\rangle\!\langle{#2}\right|}

\newcommand{\HH}{\ensuremath{\mathcal{H}}}
\newcommand{\Hf}{\ensuremath{\mathfrak{H}}}
\newcommand{\EE}{\ensuremath{\mathcal{E}}}

\newcommand{\TT}{\ensuremath{\mathcal{T}}}

\newcommand{\GG}{\ensuremath{\hat{G}}}
\newcommand{\XX}{\ensuremath{\mathcal{X}}}
\renewcommand{\SS}{\ensuremath{\mathcal{S}}}
\renewcommand{\vr}{\ensuremath{\varrho}}

\begin{document}

\title{The geometric measure of entanglement for symmetric states}


\begin{abstract}
Is the closest product state to a symmetric entangled multiparticle state also symmetric? This question has appeared in the recent literature concerning the geometric measure of entanglement. First, we show that a positive answer can be derived from results concerning symmetric multilinear forms and homogeneous polynomials, implying that the closest product state can be chosen to be symmetric. We then prove the stronger result that the closest product state to any symmetric multiparticle quantum state is \emph{necessarily} symmetric. Moreover, we discuss generalizations of our result and the case of
translationally invariant states, which can occur in spin models.
\end{abstract}

\author{Robert H\"ubener}
\affiliation{Institut f\"ur Theoretische Physik,
Universit\"at Innsbruck, Technikerstra{\ss}e 25,
6020 Innsbruck, Austria}

\author{Matthias Kleinmann}
\affiliation{Institut f\"ur Quantenoptik und Quanteninformation,
~\"Osterreichische Akademie der Wissenschaften,
Otto Hittmair-Platz 1,
6020 Innsbruck, Austria}

\author{Tzu-Chieh Wei}
\affiliation{Institute for Quantum Computing and Department of 
Physics and Astronomy, University of Waterloo, Waterloo
N2L 3G1, Canada}

\author{Carlos Gonz\'alez-Guill\'en}
\affiliation{Dpto.\ An\'alisis Matem\'atico \& IMI, Universidad Complutense de Madrid, 28040 Madrid, Spain}

\author{Otfried G\"uhne}
\affiliation{Institut f\"ur Quantenoptik und Quanteninformation,
~\"Osterreichische Akademie der Wissenschaften,
Otto Hittmair-Platz 1,
6020 Innsbruck, Austria}
\affiliation{Institut f\"ur Theoretische Physik,
Universit\"at Innsbruck, Technikerstra{\ss}e 25,
6020 Innsbruck, Austria}



\maketitle

\section{Introduction}

Entanglement is a key phenomenon in quantum mechanics and its quantification is vital for the field of quantum information theory. Many entanglement measures have been proposed for the two-particle as well as for the multiparticle case~\cite{plenio,witnessreview}. Virtually all of the proposed entanglement measures, however, suffer a serious drawback: They are very difficult to compute as their definition contains optimizations over certain states or quantum information protocols~\cite{concurrence}. Such optimizations can be performed successfully for special cases only, for instance, if the density matrix under investigation possesses a high symmetry or belongs to a special family, e.g., with low rank~\cite{calculations}.

An often-used entanglement measure for multiparticle systems is the geometric measure of entanglement~\cite{geo}. For a given multiparticle state $\ket{\psi}$, one first considers the closest fully separable state $\ket{\phi} = \ket{a}\ket{b}\ket{c}\cdots$ in terms of the overlap
\bee
G({\psi})=
\max_{\ket{\phi} = \ket{a}\ket{b}\ket{c}\cdots}
|\braket{\psi}{\phi}|,
\label{gdef}
\eee
and then defines the geometric measure of the pure state as 
\bee
E_G(\ket{\psi})= 1- G({\psi})^2.
\label{geo1}
\eee
Sometimes, the geometric measure for pure states is also taken as $\varepsilon_G(\ket{\psi})=-2\log_2 G(\psi)$. Based on this definition, the geometric measure is extended to mixed states via the convex roof construction: For a given density matrix $\vr$ one minimizes over all possible decompositions of $\vr$ into pure states $\vr = \sum_k p_k \ketbra{\phi_k}{\phi_k}$, where the $p_k$ form a probability distribution,
\be
E_G(\vr) &= &\min_{p_k, \ket{\phi_k}} \sum_k p_k E_G(\ket{\phi_k}).
\label{geo2}
\ee
Clearly, also this optimization is not straightforward to compute.

The geometric measure has become one of the widely used entanglement measures for the multiparticle case. It fulfills all the desired properties of an entanglement monotone~\cite{geo}. Moreover, it has a physical interpretation of quantifying the difficulty in distinguishing multiparticle quantum states by local means~\cite{hayashi}. It has also been used to study quantum phase transitions in spin models~\cite{geospin,weiextra} and the usefulness of states as resources for measurement based quantum computation~\cite{eisert}. The value of $E_G$ has been computed for many pure states~\cite{akimasa, hayashi2, tamaryan}, and the convex roof for some important cases has been calculated in Refs.~\cite{geo,gbb}.

If one considers the optimization problem in Eq.~\eqref{gdef}, a natural question arises whether for a symmetric state $\ket{\psi}$ the closest product state can be chosen symmetric, i.e., $\ket{\phi} = \ket{a}\ket{a}\ket{a}\cdots$. If this is true, it drastically simplifies the calculation of the geometric measure for pure symmetric states, as the number of parameters in this optimization then does not depend on the number of particles anymore. Recently this problem drew considerable attention in quantum information theory and some effort was made to prove it. For example, it has been used as a conjecture in Ref.~\cite{geo}. In Ref.~\cite{hayashi2} it has been proved for two particles that there is always a symmetric state which gives the maximum value (but it can happen that also non-symmetric states yield the same value) and a first attempt for the $N$-particle case was given. Quite recently, special cases of this conjecture have been verified~\cite{recent}, and related conjectures have been formulated~\cite{GPMSCZ09}.

In this paper, we investigate the conjecture from several perspectives. We show that a result on $N$-homogeneous polynomials over Banach spaces can be applied to the above problem and proves that the maximum in Eq.~\eqref{gdef} can be achieved by a symmetric state. However, this result does not allow to conclude that only symmetric solutions exist. We then go on to show that the optimal state maximizing $G({\psi})$ is {\it necessarily} symmetric for three or more particles. Finally, we will discuss consequences and generalizations of our results, concerning, among others, the maximization of the expectation value of symmetric positive operators.

\section{The main result}\label{smr}

In this section we will first apply a result from the theory of homogeneous polynomials to the conjecture, proving that a symmetric state attains the maximum. Then we proof a stronger version, stating that the maximizing state is necessarily symmetric. Let us fix our notation. We write the overlap with a product state as an evaluation of a corresponding $N$-linear form, i.e.,
\bee
\braket{\psi}{\alpha_1,\alpha_2,\dotsc, \alpha_N} =: 
\psi( \alpha_1, \alpha_2, \dotsc, \alpha_N).
\eee
A symmetric state $\ket\psi$ then corresponds to a symmetric $N$-linear form $\psi$. For a $k$-level system, the vectors $\alpha_i$ are elements of $\mathbbm{C}^k$ and we will always assume they are normalized. Furthermore, the expression $\alpha \equiv \beta$ denotes equality up to a phase, i.e., there is a phase $\varphi$ such that $\alpha = e^{i \varphi} \beta.$

For any symmetric $N$-linear form there exists an associated $N$-homogeneous polynomial $\hat\psi$ \cite{hopo} by virtue of the mapping
\begin{equation}
 \hat\psi\colon \alpha\mapsto \psi(\alpha,\dotsc,\alpha).
\end{equation}
The norm of this polynomial is defined by
\begin{equation}
 \lVert \hat\psi \rVert= \max_\alpha \lvert \hat\psi(\alpha) \rvert.
\end{equation}
In this context the quantity $G(\psi)$ we introduced above is denoted by $\lVert\psi \rVert$.

The relation between the norms $\lVert\hat\psi\rVert$ and $\lVert\psi\rVert$ is studied in the theory of polynomials over Banach spaces, cf.\ Ref.~\cite{Dineen}. For a Banach space $\EE$ the \emph{polarization constant} $c(N,\EE)$ is the smallest positive number such that
\begin{equation}
 \lVert \psi \rVert \leq c(N,\EE) \lVert \hat \psi \rVert
\end{equation}
holds for every symmetric $N$-linear form $\psi$ over $\EE$.

For finite dimensional real and complex Hilbert spaces this polarization constant is known to be $1$, cf.\ Ref.~\cite{kellogg,hoermander}. An exhaustive discussion can be found in Ref.~\cite{Dineen}. In our particular case $\EE=\mathbbm C^k$ it follows that symmetric product states maximize the overlap with any symmetric state, but leaves open if also non-symmetric states attain the maximal value. This is in fact never the case for $N \geq 3$, as we summarize in

\begin{lemma}\label{l:main}
Let $\psi \neq 0$ be a symmetric $N$-linear form over $\mathbbm{C}^k$ with $N \geq 3$ and let the vectors $\alpha_1, \dotsc, \alpha_N$ maximize $|\psi|$, i.e.,
\bee
G(\psi) = | \psi(\alpha_1, \alpha_2,\dotsc,\alpha_N) |.
\eee
Then the vectors $\alpha_k$ are equal up to a phase, in other words, the span of $\alpha_1, \dotsc, \alpha_N$ is one-dimensional.
\end{lemma}

In order to prove Lemma~\ref{l:main}, we will first consider the situation where $N=2.$ The following Lemma and its proof were already given Ref.~\cite{hayashi2}, however, the proof provides some observations that are essential in order to establish our main result.

\begin{lemma}[\cite{hayashi2}]\label{l:hayashi}
For any symmetric two-linear form $\psi$ over $\mathbbm{C}^k$ we can find a vector $\alpha$ such that
\bee
G(\psi) = | \psi(\alpha, \alpha) |.
\eee
\end{lemma}

Rephrasing the statement of the Lemma, when maximizing $|\psi|$ for two particles, the maximum can by reached by a symmetric choice of vectors, although a solution maximizing $|\psi|$ is not necessarily symmetric.

\begin{proof}[Proof of Lemma~\ref{l:hayashi}.]
In a fixed orthonormal basis $\{b_i\}$ the symmetric quadratic form $\psi$ is represented by a symmetric matrix $\Psi_{ij}:= \psi(b_i,b_j)$. Then we have $\psi(\alpha ,\beta) = \alpha^T \Psi \beta$, where on the right-hand side $\alpha$ and $\beta$ are column vectors with coefficients in the basis $\{b_i\}$. For a complex symmetric matrix $\Psi = \Psi^T$ Takagi's factorization theorem~\cite{takagi} states that $\Psi$ can be written as
\bee
\Psi = U^T D U,
\eee
with a unitary matrix $U$ and a diagonal matrix $D=\mathrm{diag}(r_1,\dotsc,r_{k})$ where the non-negative values $r_i$ are in decreasing order, $r_1 \ge r_2 \ge \cdots \ge r_{k} \ge 0$. In this form, a symmetric choice of $\alpha$ and $\beta$ maximizing $|\psi|$ becomes evident, namely $\alpha \equiv \beta \equiv U^{\dagger} e_1$ where $e_1=(1,0,\dotsc,0)^T$. Hence $G(\psi)=r_1$.
\end{proof}

Let us make some remarks on the remaining freedom in the choice of $\alpha$ and $\beta.$ First, we note that if $r_1 > r_2$ the only choice to reach the maximum $r_1$ is the one given in the proof above. Hence for this case the maximizing solution is unique (up to a phase) and symmetric.

Otherwise, consider the case $r_1 = r_2 = \cdots = r_d$. We then say $\psi$ is degenerate and define $R_1 := \mathrm{span}_{\mathbbm{C}}(\{e_1, \dotsc, e_d\})$. If $G(\psi)= | \psi(\alpha, \beta) |$, we can always write $\alpha \equiv U^{\dagger} e^* $ and $\beta \equiv U^{\dagger} e$ with some $e \in R_1$, where the vector $e^* \in R_1$ denotes the vector obtained from $e$ by complex conjugation in the given basis. The case that $e \equiv e^*$ then corresponds to the symmetric solutions.

The symmetric maximizing solutions therefore correspond via $U$ to real vectors in $R_1$ (up to a phase), and having a non-symmetric solution $\alpha \not\equiv \beta$ of the maximization implies degeneracy. Moreover, in case of degeneracy we find a continuum of inequivalent asymmetric as well as symmetric solutions. The following observation will be needed in the main proof and expresses this fact.

\begin{observation}\label{o:bservation}
Given a symmetric two-linear form $\psi$ with
\bee
G( \psi ) = | \psi(\alpha, \beta) |,
\eee
where $\beta \not\equiv \alpha$, we can always find two orthonormal vectors $\delta_1$ and $\delta_2$, such that:
\begin{itemize}
\item[(i)] $\delta_1$ and $\delta_2$ span the same space as $\alpha$ and $\beta$, in particular $\alpha, \beta \in \mathrm{span}_{\mathbbm{C}}(\{\delta_1, \delta_2\}).$
\item[(ii)] We have $G(\psi) = |\psi(\delta_1, \delta_1)| = |\psi(\delta_2, \delta_2)| = |\psi(\eta, \eta)| = |\psi(\mu, \mu')|$, where
\bee
\begin{split}
\eta &:= (\delta_1 + \delta_2)/\sqrt{2},\\
\mu &:= (\delta_1 + i \delta_2)/\sqrt{2},\\
\mu' &:= (\delta_1 - i \delta_2)/\sqrt{2}.
\end{split}
\eee
\item[(iii)] The vectors $\delta_1, \delta_2, \eta, \mu'$ do not equal $\alpha$, even modulo a phase.
\end{itemize}
\end{observation}

\begin{proof}
Consider $G(\psi) = |\psi(\alpha, \beta)| = |\psi(U^{\dagger} e^*, U^{\dagger} e)|$ with the unitary matrix $U$ from Takagi's factorization of $\Psi$, where $e \not\equiv e^*$ as $\beta \not\equiv \alpha$ by assumption. We can choose two real orthonormal vectors $f_1, f_2$ such that 
$
\mathrm{span}_{\mathbbm{C}}(\{f_1,f_2\}) = \mathrm{span}_{\mathbbm{C}}(\{e,e^*\}).
$ 
They can be obtained from the real and imaginary parts of $e$ and $e^*$~\cite{eestar}. Hence
\bee
\delta_1 = U^{\dagger} f_1 \mbox{ and } \delta_2 = U^{\dagger} f_2
\eee
is a valid choice of orthonormal vectors fulfilling (i), each providing a symmetric maximization of $|\psi|$. The vectors $\delta_1,\delta_2$ as well as the vectors $\eta,\mu, \mu'$ derived from $\delta_1,\delta_2$ fulfill (ii) by construction, according to the observations above and because $\eta = U^{\dagger}(f_1 + f_2)/\sqrt{2}$, where $(f_1 + f_2)/\sqrt{2} \in R_1$ (likewise for $\mu, \mu'$). Finally, we have enough freedom to choose $f_1, f_2$ to satisfy (iii).
\end{proof}

After these preliminaries, we are ready to prove Lemma~\ref{l:main}:

\begin{proof}[Proof of Lemma~\ref{l:main}]
The proof consists of two parts. In Part I we prove the case $N=3$. In Part II we extend the result to arbitrary $N > 3$.

\emph{Part I.} Assume a maximizing set of vectors $\{ \alpha, \beta, \gamma \}$ has been found,
\bee
G(\psi)=|\psi(\alpha, \beta, \gamma)|.
\eee
We show that the assumption $\mathrm{dim}[\mathrm{span}_{\mathbbm{C}}(\{ \alpha, \beta, \gamma \})] \neq 1$ leads to $\psi=0$.

Without losing generality, we hence assume that $\gamma \not\equiv \beta$. Then we have a degenerate quadratic form $\psi(\alpha, \cdot, \cdot \cdot)$. Using Lemma~\ref{l:hayashi} we obtain a symmetric maximizing solution $\sigma \not\equiv \alpha$ (due to $\gamma \not\equiv \beta$, $\sigma$ is not unique). Hence we have $G(\psi)=|\psi(\alpha, \sigma, \sigma)|$. {From} Observation~\ref{o:bservation}, applied to the quadratic form $\psi(\cdot, \cdot \cdot, \sigma)$, we take the vectors $\delta_1, \delta_2, \eta, \mu, \mu'$ with the properties as stated. With these vectors we define the $2 \times 2$-matrices $A,B, N, M$ via
\bee
\begin{split}
\label{matdef}
A_{kl} &:= \psi(\delta_1, \delta_k, \delta_l),\\
B_{kl} &:= \psi(\delta_2, \delta_k, \delta_l),\\
N_{kl} &:= \psi(\eta, \delta_k, \delta_l) = (A_{kl} + B_{kl})/\sqrt{2},\\
M_{kl} &:= \psi(\mu, \delta_k, \delta_l) = (A_{kl} + i B_{kl})/\sqrt{2}.
\end{split}
\eee
As $\sigma,\delta_1, \delta_2, \eta$, and $\mu'$ are in $\mathrm{span}_{\mathbbm{C}}(\{\delta_1, \delta_2\})$ the matrices $A,B,N,M$ correspond to two-forms which assume the maximum $r_1=G(\psi)$ on this span. Since $\delta_1, \delta_2, \eta, \mu' \not\equiv \sigma$, the quadratic forms $A,B,N,M$ are degenerate with the value $r_1$. In terms of the Takagi factorization, we can write $A=U^T_A D U^{\phantom{A}}_A$ and $B=U^T_B D U^{\phantom{B}}_B$, etc.\ with $D=\mathrm{diag}(r_1, r_1).$ It follows that $A^{\dagger} A = B^{\dagger} B = N^{\dagger} N = M^{\dagger} M = D^2$, which implies $A^\dagger B +B^\dagger A = i(A^\dagger B-B^\dagger A) = 0$. Hence $B^{\dagger} A = 0$, which is only possible for $\psi = 0$.

\emph{Part II.} The extension to the case $N>3$ is proved as follows. If $G(\psi) = |\psi(\alpha_1,\alpha_2,\dotsc,\alpha_N)|$ we define a symmetric 3-linear form by
\bee
\psi_{\hat{k}} := \psi(\hat{\alpha}_1,\hat{\alpha}_2,\dotsc,\hat{\alpha}_k,\dotsc,\alpha_N)
\eee
where $\hat{\alpha}_i$ denotes omission, i.e., $\alpha_1,\alpha_2,\alpha_k$ are omitted. As in Part 1, we have $G(\psi)=|\psi_{\hat{k}}(\alpha_1,\alpha_2,\alpha_k)|$ and thus $\alpha_1 \equiv \alpha_2 \equiv \alpha_k$ for all $k$. Hence all vectors $\alpha_1, \dotsc, \alpha_N$ must be the same up to a phase.
\end{proof}

Lemma~\ref{l:main} is stated for $N$-linear forms over a {complex} vector space. If $\ket{\psi}$ is real, one can also consider the maximization over real product vectors. In general this will yield a different result than the complex case \cite{realremark}. Since the polarization constant is $1$ also in the real case, one can find a a symmetric state  among the real product states which attains the maximum. In contrast to the complex case, the maximizing state is, however, not necessarily symmetric for three particles. A counterexample is $\ket\psi=(\ket{001}+\ket{010}+\ket{100}-\ket{111})/2$ where the maximum of $1/2$ is also attained by $\ket\phi=\ket{001}$.

The following Lemma provides an additional very simple proof that $c(N,\HH)=1$ for $N=2^\ell$, and is based on a symmetrization procedure.

\begin{lemma}\label{l:simple}
Let $\psi$ be a symmetric $N$-linear form with $N=2^\ell$ and let the vectors $\alpha_1,\dotsc,\alpha_N$ maximize $|\psi|$. Then there exists a normalized vector $\zeta$ in the span of $\alpha_1,\dotsc,\alpha_N$ such that
\begin{equation}
|\psi(\alpha_1,\dotsc,\alpha_N)|= |\psi(\zeta,\dotsc,\zeta)|.
\end{equation}
This statement holds for real and complex Hilbert spaces.
\end{lemma}

\begin{proof}
Let $\langle\cdot,\cdot\cdot\rangle$ denote the scalar product and we define the constant $\Lambda=\psi(\alpha_1,\alpha_2,\alpha_3,\dotsc)$. Then $\psi(\cdot,\alpha_2,\alpha_3,\dotsc)= \Lambda\, \langle \alpha_1 , \cdot \rangle$ and $\psi(\alpha_1,\,\cdot\,,\alpha_3,\dotsc)= \Lambda\, \langle \alpha_2 , \,\cdot\, \rangle$ (cf.\ Eq.~(6) in Ref.~\cite{geo}). Using the symmetry and linearity of $\psi$ we arrive at
\be
\psi(\cdot,\beta_1,\alpha_3,\dotsc)=\Lambda\, \langle \beta_1, \,\cdot\, \rangle
\ee
and hence
\be
\Lambda= \psi(\beta_1,\beta_1,\alpha_3,\dotsc),
\ee
where $\beta_1=(\alpha_1+\alpha_2)/||\alpha_1+\alpha_2||$. (If $\alpha_1=-\alpha_2$, we set $\beta_1 = \alpha_1$ and replace $\Lambda$ by $-\Lambda$.)

We now repeat this procedure, first yielding $\Lambda =  \psi(\beta_1,\beta_1,\beta_3,\beta_3,\dotsc)$ and then $ \Lambda =  \psi(\gamma_1,\gamma_1,\gamma_1,\gamma_1,\dotsc) $, where $\beta_3$ is defined analogously to $\beta_1$ and $\gamma_1=(\beta_1+\beta_3)/||\beta_1+\beta_3||$. In the second step we applied the symmetrization to the first and third argument as well as to the second and fourth argument. Since $N=2^\ell$, we can complete this symmetrization and arrive at $ \Lambda =  \psi(\zeta,\dotsc,\zeta)$ for some $\zeta$ in the span of $\alpha_1,\dotsc,\alpha_N$.
\end{proof}

\section{Discussion}

\subsection{Physical interpretation of the proof}

An interpretation of the proof of Lemma~\ref{l:main} in physical terms is the following. The matrices $A,B,N$ and $M$ in Eq.~\eqref{matdef} are representations of the state $\ket{\psi}$, after one site has been measured out and the remaining state has been projected onto a two-dimensional subspace. The values $r_i$ correspond to Schmidt coefficients of this remaining state and, as they are equal, the state corresponds to a Bell state. The proof of Lemma~\ref{l:main} shows that for qubits it is impossible to create a state of three particles that is both symmetric and always results in a Bell-pair like state after an arbitrary measurement on one site.

\subsection{Translationally invariant states}

It is interesting to ask whether also for translationally invariant states the maximum is attained in a symmetric state, as such states occur naturally in the analysis of spin models. This has sometimes been assumed when investigating the geometric measure in condensed matter systems.

First, a counterexample for this conjecture is the state
\bee
\ket{\psi}= \frac{1}{\sqrt{2}}(\ket{0101} + \ket{1010})
\eee
for which the closest separable states are the non-symmetric states $\ket{0101}$ and $\ket{1010}.$ In fact, one can find translationally invariant states which are orthogonal to any symmetric product state, e.g., $\ket{\psi}\sim (\ket{0101}- \ket{0011}+ \mbox{ all translations}).$

This situation gets worse as the number of particles increases. Let $\TT$ denote the subspace of translationally invariant states for $N$ qukits and let $\SS \subset \TT$ be the permutationally symmetric subspace. Then any state in $\XX=\TT \cap \SS^\perp$ -- the orthocomplement of $\SS$ in $\TT$ -- has a vanishing overlap with any symmetric product state, hence the closest product state is not symmetric. The dimension of $\TT$ is given by~\cite{Polya:1937AM}
\be
\dim(\TT)= \frac{1}{N} \sum_{j\mid N} \varphi(j)\, k^{N/j},
\ee
where $\varphi$ denotes Euler's totient function and 
the summation is over all divisors $j$ of $N$. For $\SS$ we have
\be
\dim(\SS) = \begin{pmatrix}N+k-1\\k-1\end{pmatrix}.
\ee
Therefore, if $N \gg k$ then the dimension of the subspace $\XX$ is roughly given by $(k^N-N^k)/N$ and the fraction of states where the conjecture holds shrinks rapidly as the number of particles increases.

Concerning the analysis of entanglement in spin models, this shows that the assumption that the closest separable state to the ground state is symmetric, has to be handled with care. For some models, it seems to be true~\cite{weidiss,weiextra}, for other models (like the Majumdar-Ghosh model~\cite{MajGosh}) one can directly check that it is wrong.

\subsection{Operators of higher rank}

We now consider generalizations of our results. Let $\Pi_\SS$ be the projector onto the symmetric subspace $\SS$. An operator $A$ is \emph{permutationally symmetric} if it acts on the symmetric subspace only, i.e., it fulfills $A = \Pi_\SS A \Pi_\SS$. $A$ is called \emph{permutationally invariant} if it is invariant under permutation of the particles (the latter is a weaker condition than the former~\cite{gt09}). We hence define for an observable $X$ 
\bee
\begin{split}
\hat G(X)&:= \max_{\ket{\varphi}=\ket{a}\ket{b}\ket{c}\cdots}|\bra{\varphi}X\ket{\varphi}|
\\
\hat G_\SS(X)&:=\max_{\ket{\varphi}=\ket{a}\cdots\ket{a}}|\bra{\varphi}X\ket{\varphi}|
\end{split}
\eee
Such optimizations occur naturally in the construction of entanglement witnesses or in the estimation of entanglement measures via Legendre transforms~\cite{witnessreview}. 

To study the relation of these quantities, we can write $\bra{\varphi}X\ket{\varphi}=\Tr{X \ketbrad{\varphi}}$ as an evaluation of a corresponding $N$-linear form $\xi$ over $\Hf^k$ (the Banach space of Hermitian matrices of dimension $k$ equipped with the trace norm) due to
\begin{equation}
\xi(A_1,\dotsc,A_N)=\Tr{X A_1\otimes\cdots\otimes A_N}.
\end{equation}
Then any permutationally invariant operator $X$ corresponds to a symmetric $N$-linear form $\xi$ and an $N$-homogeneous polynomial $\hat\xi$. We now define $\lVert \xi \rVert$ and $\lVert \hat\xi \rVert$ analogously to Section~\ref{smr} (using $\lVert A\rVert= 1$ as normalization condition). It is straightforward to see that $\hat G(X)=\lVert \xi \rVert$ and $\hat G_\SS(X)=\lVert \hat\xi\rVert$. As the polarization constant can be shown to be \cite{pcb}
\begin{equation}
 c(N,\Hf^k)=N^N/N! \quad\text{for}\quad N\le k,
\end{equation}
the quotient $\hat G(X)/ \hat{G}_\SS(X)$ can get arbitrarily large as $N$ and $k$ increase. 

At the end of this section, we will provide a further explicit example. Let us first discuss some cases where symmetry assumptions do hold:

\begin{corollary}\label{c:operators}
(i) If X is a positive permutationally symmetric observable then $\hat G(X)$ can be attained by a symmetric state.
\\
(ii) If $X$ is a permutationally invariant $N$-qubit observable that contains only full correlation terms, then $\hat G(X)$ can be attained by a symmetric state.
\end{corollary}

\begin{proof}
(i) We note that 
\be
\hat G(X) \leq \max_{\ket{\psi} = \ket{b_1}\cdots \ket{b_n}}\max_{\ket{\varphi}=\ket{a_1}\cdots\ket{a_n}}| \bra{\varphi}X\ket{ \psi}|.
\ee
Fixing $\ket{\psi},$  the (unnormalized) state $X\ket{\psi}= \Pi_\SS X \Pi_\SS\ket\psi$ is symmetric and by virtue of Lemma~\ref{l:main} the maximum is reached by a symmetric state $\ket{a,\dotsc,a}$. Repeating the reasoning with the fixed state $\ket\varphi=\ket{a,\dotsc,a}$, we get 
$
\hat G(X) \leq \max_{\ket{b}}\max_{\ket{a}}| \bra{a,\dotsc,a}X \ket{b,\dotsc,b}|.
$
The fact that for positive operators $2 |\bra{\alpha} P\ket {\beta}| \leq \bra{\alpha} P \ket{\alpha} + \bra{\beta} P\ket {\beta} \leq 2 \max\{ \bra{\alpha} P \ket{\alpha}, \bra{\beta} P\ket {\beta}\}$ holds for arbitrary $\ket{\alpha}$ and $\ket{\beta}$ proves that $\hat G(X) \leq \hat G_\SS(X),$ hence $\hat G(X) = \hat G_\SS(X).$

(ii) An $N$-qubit operator $X$ contains only full correlation terms if $X=\sum_{i,j,\dotsc \in \{x,y,z\} } \lambda_{ij\cdots} \sigma_i \otimes \sigma_j \otimes \cdots$. Note that here $\sigma_0=\one$ does not occur; a physically relevant and well known example for such an operator $X$ is the Mermin inequality. As $X$ is permutationally invariant, $\lambda$ is equivalent to a symmetric $N$-linear form over $\mathbbm{R}^3$. The Bloch representation for qubits implies that here the maximization is equivalent to finding $\max_{r_i \in \mathbbm{R}^3} |\lambda(r_1,r_2,\dotsc,r_{N})|$ where the vectors $r_i$ are the corresponding Bloch vectors. Since the polarization constant for real Hilbert spaces is $1$ \cite{Dineen,kellogg,hoermander}, the assertion follows.
\end{proof}

Let us conclude with some examples where symmetry assumptions do not hold.
If $X$ is symmetric but not positive, then the maximum $\GG(X)$ is, in general, not attained by a symmetric state. A counterexample for two qubits is $X=6\ketbra{\psi^+}{\psi^+}-\ketbra{00}{00}-2\ketbra{11}{11}$ with $\ket{\psi^+}=(\ket{01}+\ket{10})/\sqrt{2}.$ Then $\GG(X)=3$ (we can take $\ket{\phi_0}= \ket{01}$) while the maximum for symmetric product states is 34/15. Also, if $X$ is permutationally invariant (and even positive) then the maximum $\GG(X)$ is, in general, not attained by a symmetric state. A counterexample is the singlet state, $X=\ketbra{\psi^-}{\psi^-}$ with $\ket{\psi^-}=(\ket{01}-\ket{10})/\sqrt{2}.$ This operator is invariant under permutation of the particles, but it does not act on the symmetric space. It has $\GG(X)=1/2$, but restriction to the symmetric $\ket{\phi}$ would yield again $\GG(X)=0$. This clarifies some questions raised in Ref.~\cite{GPMSCZ09}.

\section{Conclusion}
In conclusion, we have discussed a widely used conjecture concerning the geometric measure of entanglement. Our results not only simplify the calculation of the geometric measure for symmetric states, but they also have applications to subjects in condensed matter physics. Furthermore, from a mathematical perspective, the quantity $G$, as defined in the introduction, is known as the {injective tensor norm}~\cite{DF93}. On the one hand, this norm is of central importance in tensor analysis, since there are, as Grothendieck showed, fourteen inequivalent natural tensor norms derivable from $G$~\cite{grothendieck}. On the other hand, this norm has also occurred in the discussion of the maximal output purity of quantum channels \cite{WHP}. So we believe that the study of tensor norms can yield further interesting insights in quantum information theory.

We thank A.~Abdesselam, A.~Harrow, M.~Murao, R.~Orus, D. P\'erez-Garc\'ia, S.~Virmani and R.F.~Werner for discussions. This work has been supported by the FWF (START prize) and the EU (OLAQUI, QICS, SCALA). TCW acknowledges support from IQC, NSERC and ORF. CGG acknowledges financial support from the Spanish grants I-MATH, MTM2008-01366, CCG08-UCM/ESP-4394 and Beca-COMPLUTENSE2006.


\begin{thebibliography}{99}
\bibitem{eestar}
Note that $e$ and $e^*$ are vectors with complex conjugated elements, i.e., $e= e_r + i e_i$ and $e^* = e_r - i e_i$ where $e_r,e_i \in \mathbbm{R}^k$. Hence $e_r,e_i$ is an example for a real basis, being linearly independent since $e \not\equiv e^*$. In fact, any real orthonormal basis of $\mathrm{span}_{\mathbbm{C}} (\{e_r,e_i\})$ can be used as $f_1, f_2$, as long as they fulfill (iii).
\bibitem{realremark} 
An example of a for four-qubit state where the maximizations over real and complex product vectors differ is $\ket{\psi} = [\ket{0000}+\ket{0001} +\ket{0010}+\ket{0100}+\ket{1000} -2(\ket{1110}+\ket{1101}+\ket{1011} +\ket{0111})]/\sqrt{21}.$
\bibitem{pcb}
One can embed isometrically $l_1^k$ (i.e., $\mathbbm{R}^k$ with norm $\Vert \cdot \Vert_1$) into $\Hf^k$ by a direct mapping into the diagonal. As $c(N,l_1^k)=N^N/N!$ for $N\le k$ \cite{Dineen} and on the other hand $c(N,\EE)\le N^N/N!$ for any Banach space $\EE$, we have $c(N,\Hf^k)=N^N/N!$ for $N\le k$.
\end{thebibliography}
\end{document}